\documentclass[envcountsect, envcountsame]{llncs}

\usepackage{textcomp}
\usepackage{amsmath}
\usepackage{amssymb}
\usepackage{graphics}
\usepackage{psfrag}

\newcommand{\vbl}{\ensuremath{{\rm vbl}}}

\newcommand{\poly}{\ensuremath{{\rm poly}}}

\begin{document}
\frontmatter
\pagestyle{headings}
\title{Satisfiability of Almost Disjoint CNF Formulas}

\author{Dominik Scheder}
\institute{Theoretical Computer Science, ETH Z\"urich\\
CH-8092 Z\"urich, Switzerland\\
\email{dscheder@inf.ethz.ch}}
\maketitle
\begin{center}
 \today
\end{center}

\begin{abstract}
  We call a CNF formula {\em linear} if any two clauses have at most
  one variable in common. Let $m(k)$ be the largest integer $m$ such that
  any linear $k$-CNF formula with $\leq m$ clauses is satisfiable.
  We show that $\frac{4^k}{4e^2k^3} \leq m(k) < \ln(2) k^4 4^k$.

  More generally, a $(k,d)$-CSP is a constraint satisfaction problem
  in conjunctive normal form where each variable can take on one of $d$ 
  values, and each constraint contains $k$ variables and 
  forbids exacty one of the $d^k$ possible assignments to these variables.
  Call a $(k,d)$-CSP $\ell$-disjoint if no two distinct constraints have
  $\ell$ or more variables in common. Let $m_{\ell}(k,d)$ denote the
  largest integer $m$ such that any $\ell$-disjoint $(k,d)$-CSP with	
  at most $m$ constraints is 
  satisfiable. We show that

  $$\frac{1}{k} \left(\frac{d^k}{ed^{\ell -1}k}\right)^{1+\frac{1}{\ell -1}}
  \leq m_{\ell}(k,d) < c\left(k^2\ell^{-1}\ln(d)d^k\right)^{1+\frac{1}{\ell -1}} \ .
  $$
  for some constant $c$.  This means for constant $\ell$, upper and
  lower bound differ only in a polynomial factor in $d$ and $k$.
\end{abstract}

\section{Introduction}

How difficult is it to come up with an unsatisfiable CNF formula?
Stupid question, you might think: $\{\{x\},\{\bar{x}\}\}$, here is
one. Two clauses, each containing one literal, and unsatisfiable.
Well, yes, but what if we want a $k$-CNF formula, i.e., we require
that every clause contains exactly $k$ literals? Now it's a little bit
less trivial, but still easy: Take a clause $\{x_1,x_2,\dots,x_k\}$,
then $\{\bar{x}_1,x_2,\dots,x_k\}$, $\{x_1,\bar{x}_2,\dots,x_k\}$,
until you have exhausted all $2^k$ combinations of negative and
positive literals. Each assignment to the $k$ variables is ruled out
by exactly one clause: Your formula has $2^k$ clauses, and it is
unsatisfiable. This formula is the ``simplest'' unsatisfiable $k$-CNF
formula, in a sense as $K_{k+1}$ is the simplest non-$k$-colorable
graph. What if we impose further restrictions? For example, what if no
variable can occur in more than one clause? This restriction is surely
too strong: One can satisfy each clause individually, hence such
a formula is always satisfiable, unless it contains the empty clause.\\

Let us consider two weaker restrictions. First, what if each variable
may occur in several clauses of our $k$-CNF formula, but in at most
$d$? Let us call such a formula a {\em $d$-bounded $k$-CNF formula}.
Second, what if we allow every {\em pair} of variables to occur in at
most one clause, or, equivalently, allow any two clauses to have at
most one variable in common? Such a formula is called, in analogy to
hypergraph terminology, a {\em linear} $k$-CNF formula.\\

The first problem has been introduced by Tovey~\cite{Tovey84}, who
showed, using Hall's Marriage Theorem, that every $k$-bounded $k$-CNF
formula is satisfiable. This has been improved by Kratochv\'{i}l,
Savick\'{y} and Tuza~\cite{KST93}, who proved that there is some
threshold function $f(k)$ such that any $f(k)$-bounded $k$-CNF formula
is satisfiable, but deciding satisfiability of $f(k)+1$-bounded
$k$-CNF formulas is already NP-complete, and further, that $f(k) \geq
\frac{2^k}{ek}$.  For an upper bound on how often we can allow a
variable to occur while still guaranteeing satisfiability, Hoory and
Szeider~\cite{HS06} show how to construct unsatisfiable $d$-bounded
$k$-CNF formula for $d \in
mathcal{O}\left(\frac{\ln(k)2^k}{k}\right)$-CNF formulas.
Thus, $f(k)$ is known up to a logarithmic factor.\\

For the second question, let us give an unsatisfiable linear
$2$-CNF formula: 
$$
\{ \{\bar{u},v\},\{\bar{v},w\}, \{\bar{w},x\}, \{\bar{x},u\},
\{u,w\},\{\bar{v},\bar{x}\} \} \ .
$$
This formula has $6$ clauses, which is as few as possible for
unsatisfiable linear $2$-CNF formulas. Finding an unsatisfiable
$3$-CNF formula is already much harder. Hence we may ask the following
question: For which $k$ do unsatisfiable linear $k$-CNF formulas
exist, and if they exist, how many clauses do they have?  The
existence question has been answered by Porschen, Speckenmeyer and
Zhao~\cite{PSZ08}, who give an explicit construction of unsatisfiable
linear $k$-CNF formulas, for any $k$.  However, the size of their
formulas (i.e., the number of clauses), is gigantic: Let $m(k)$ be the
size of the unsatisfiable linear $k$-CNF formula obtained by the
construction in ~\cite{PSZ08}.  Then $m(0) = 1$ and $m(k+1) =
m(k)2^{m(k)}$. In this paper, we prove that much smaller unsatisfiable
linear $k$-CNF formulas exist, namely of size $\poly(k)4^k$, and
complement this by proving a lower bound of $\frac{4^k}{\poly(k)}$.
Since the smallest non-linear unsatisfiable $k$-CNF formula has
exactly $2^k$ clauses, this shows that unsatisfiable linearity
formuals require
significantly more clauses than non-linear ones.\\

A similar problem has been investigated, and to large extent solved,
for hypergraphs: An $r$-hypergraph $\mathcal{H}$ is a hypergraph where
every edge has $r$ vertices, and a proper $k$-coloring of
$\mathcal{H}$ is a coloring of the vertices such that no edge is
monochromatic. A hypergraph is called {\em linear} if $|e_1 \cap e_2|
\leq 1$ for any two distinct edges $e_1, e_2$ of $\mathcal{H}$. It is
easy to construct a non-$k$-colorable $r$-hypergraph, for any $k$ and
$r$.  However, it is not obvious whether non-$k$-colorable {\em
  linear} $r$-hypergraphs exist. For $k=2$, this has been positively
answered by Abbott~\cite{Abbott65}. For general $k$, existence follows
from the Hales-Jewett theorem~\cite{HJ63}.  Using Ramsey-like
theorems, the obtained bounds on the size of $\mathcal{H}$ have been
quite poor.  Tight bounds --- up to a constant factor --- have later
been given by Kostochka, Mubayi, R\"odl and
Tetali~\cite{KMRT01}, usign probabilistic techniques.\\

\subsection{Notation and Terminology}

Though we are primarily interested in linear $k$-CNF formulas, our
methods apply to a much more general class, namely $(k,d)$-constraint
satisfaction problems, or short $(k,d)$-CSPs. This is basically the
same as a $k$-CNF formula, only that each variable can take on one
of $d$ different values, not just $2$ as in the binary case.  In this
context, a literal is an inequality $x \ne b$, where $x$ is a variable
and $b \in \{0,1,\dots,d-1\}$. A $k$-constraint is a set of set of $k$
literals, and a $(k,d)$-CSP is a set of $k$-constraints.  An {\em
  assignment} is a mapping from variables to $\{0,1,\dots,d-1\}$.  An
assignments $\alpha$ {\em satisfies} a literal $x \ne b$ if, well,
$\alpha(x) \ne b$. It satisfies a constraint if it satisfies {\em at
  least one} literal in it, and it satisfies a CSP if it satisfies
every constraint of it. An issue that sometimes causes confusion is
whether one allows a constraint to contain several literals involving
the same variable. We do not. However, this is not important, since
such a constraint, e.g., $\{x \ne 0, x \ne 1\}$ would be satisfied by
every assignment anyway. \\

We say variable $x$ {\em occurs} in constraint $C$ if $C$ contains the
literal $x \ne b$ for some $b \in \{0,1,\dots,d-1\}$. For a CSP $F$,
we denote by $\deg(x,F)$ the number of constraints $C \in F$ in which
$x$ occurs, and by $\vbl(C)$ the set of all variables occurring in
constraint $C$. For example $\vbl( \{x \ne 0, y \ne 1, z \ne 1\}) =
\{x,y,z\}$. A CSP $F$ is called {\em $\ell$-disjoint} if there are no
two distinct constraints $C, D \in F$ with $|\vbl(C) \cap \vbl(D)|
\geq \ell$.  Thus, a linear $k$-CNF formula is a $2$-disjoint
$(k,2)$-CSP.

\subsection{Results}

Let $m(k)$ be the largest integer $m$ such that any linear $k$-CNF
formula with $\leq m$ clauses is satisfiable.  For CSPs, let
$m_\ell(k,d)$ denote the largest integer $m$ such that any
$\ell$-disjoint $(k,d)$-CSP with at most $m$ constraints is
satisfiable. Clearly $m_2(k,d) = m(k)$. Our main result is

\begin{theorem}
  There is some constant $c > 0$ such that
  \begin{eqnarray}
    \frac{1}{k} \left(\frac{d^k}{ed^{\ell -1}k}\right)^{1+\frac{1}{\ell -1}}
    \leq m_{\ell}(k,d) < 
    c \left(k^2\ell^{-1}\ln(d)d^k\right)^{1+\frac{1}{\ell -1}} \ .
  \end{eqnarray}
\label{main-result}
\end{theorem}

To understand these bounds, suppose $\ell$ is constant.
Then the dominating term is $d^{k(1+{\frac{1}{\ell-1}})}$ in both the
upper and lower bound, and the two bounds differ only by a polynomial
factor in $k$ and $d$.  For linear $k$-CNF formulas, we obtain

\begin{eqnarray}
  \frac{4^k}{4e^2k^3} \leq m(k) < k^4 4^k \ .
\end{eqnarray}

Compare this with the bound for general $(k,d)$-CSPs: The smallest
unsatisfiable $(k,d)$-CSP has exactly $d^k$ constraints.

\section{A Lower Bound}
\label{section-lower}

Our main tool to prove a lower bound is the symmetric version of the
Lov\'asz Local Lemma (see e.g.~\cite{AS00}):

\begin{lemma}[Lov\'asz Local Lemma]
  Let $\mathcal{E}_1,\dots,\mathcal{E}_n$ be events in a probability
  space with $\Pr[\mathcal{E}_i] \leq p$ for every $i$. If each event
  $\mathcal{E}_i$ is independent of all other events except at most
  $d$ many, and $ep(d+1) \leq 1$, then $\Pr[\bigcup \mathcal{E}_i] < 1$.
  \label{lemma-local}
\end{lemma}

The following corollary states that any CSP is satisfiable unless some
variable occurs ``too often''. This has been shown by~\cite{KST93} for
$d=2$, and their proof directly generalizes to general $d$.

\begin{corollary}
  If $F$ is a $(k,d)$-CSP and $\deg(x,F) \leq \frac{d^k}{ek}$
  for every variable $x$, then $F$ is satisfiable.
  \label{low-deg}
\end{corollary}

\begin{proof}
  Assign each variable uniformly at random a value from
  $\{0,1,\dots,d-1\}$.  Write $F = \{C_1,\dots,C_m\}$ and let
  $\mathcal{E}_i$ be the event that constraint $C_i$ is not satisfied.
  Clearly $p:= \Pr[\mathcal{E}] = d^{-k}$.  Event $\mathcal{E}_i$ is
  independent of all other events except those events $\mathcal{E}_j$
  where $\vbl(C_i) \cap \vbl(C_j) \ne \emptyset$, i.e. those
  constraints sharing a variable with $C_i$. Since $\vbl(C_i)$
  contains $k$ variables, and each occurs in at most
  $\frac{d^k}{ek}-1$ other clauses, $C_i$ shares a variable with at
  most $k\left(\frac{d^k}{ek}-1\right) \leq e^{-1}d^k-1$ other
  clauses. By Lemma~\ref{lemma-local}, with positive probability none
  of the events $\mathcal{E}_i$ occurs, i.e., $F$ is satisfiable.
  \qed
\end{proof}

Let $F$ be a $(k,d)$-CSP. We call $x$ {\em frequent in $F$} if
$\deg(x,F) > \frac{d^k}{ed^{\ell -1}k}$. Our idea is that an
$\ell$-disjoint $(k,d)$-CSP with few frequent variables can be
transformed into a $(k-\ell+1,d)$-CSP $F'$ having no frequent
variable. By Corollary~\ref{low-deg}, $F'$ is satisfiable, and the
transformation is such that $F$ is satisfiable, too.

\begin{theorem}
  Any $\ell$-disjoint $(k,d)$-CSP with $\leq \left(\frac{d^k}{ed^{\ell
        -1}k}\right)^{\frac{1}{\ell -1}}$ frequent variables is
  satisfiable.
\label{few-frequent}
\end{theorem}

\begin{proof}
  We obtain a new formula $F'$ by removing certain literals from
  certain clauses: For each constraint $C \in F$, we distinguish two
  cases: If $C$ contains less than $\ell$ variables that are frequent
  in $F$, let $C'$ by $C$ minus all literals involving one of these
  frequent variables. Otherwise, let $C'$ just be $C$.  We define $F'
  := \{C' \big| C \in F\}$. Observe that $F'$ contains constraints of
  different sizes, ranging from $k-\ell+1$ to $k$. Further, for
  each constraint in $C' \in F'$, the number of variables in 
  $\vbl(C')$ that are frequent in $F$ is either $0$ or $\geq \ell$.
  
  We claim that $\deg(x,F') \leq \frac{d^k}{ed^{\ell -1}k}$ for any
  variable $x$. If $x$ is not frequent in $F$, this is obvious, since
  $\deg(x,F')\leq \deg(x,F)$.  If $x$ is frequent in $F$, let $C_1,
  \dots, C_t$, $t:=\deg(x,F')$ be the clauses of $F'$ containing $x$.
  Clearly, each $C_i$ contains $x$, which is frequent in $F$.  For
  each $C_i \in F'$ containing $x$, $C_i$ contains at least $\ell -1$
  variables besides $x$ which are frequent in $F$. We pick $\ell-1$ of
  them arbitrarily and call this set $D_i$.  Clearly $D_i \ne D_j$ for
  $i \ne j$, otherwise the $\ell$-set $D_i \cup \{x\}$ would occur in
  $C_i$ and $C_j$, contradicting $\ell$-disjointness of $F'$.  Let $n$
  be the number of frequent variables in $F$. There are at most ${n
    \choose \ell -1}$ choices for an $(\ell-1)$-set of frequent
  variables, thus
  $$
  \deg(x,F') = t \leq {n \choose \ell -1} \leq n^{\ell -1} 
  \leq \frac{d^k}{ed^{\ell-1}k} \ .
  $$  
  We would now like to apply Corollary~\ref{low-deg} for
  $(k-\ell+1,d)$-CSPs.  However, $F'$ is not a $(k-\ell+1,d)$-CSP,
  because it may still contain larger constraints. This is no problem,
  as we can further delete literals until every constraint has size
  exactly $(k-\ell+1)$.  This process clearly does not increase any
  $\deg(x,F')$. Hence, by Corollary~\ref{low-deg}, $F'$ is
  satisfiable, and so is $F$.  \qed
\end{proof}

\textbf{Proof of the lower bound in Theorem~\ref{main-result}.}
Assume $F$ is an unsatisfiable $\ell$-disjoint $(k,d)$-CSP. Then by
Theorem~\ref{few-frequent}, we have
$\left(\frac{d^k}{ed^{\ell-1}k}\right)^{\frac{1}{\ell -1}}$ frequent
variables. Since
$$
\sum_{C\in F} |C| = \sum_x \deg(x,F)
$$
and $|C| = k$ for all $C\in F$, it follows that $F$ has more than
$\frac{1}{k} \left(\frac{d^k}{ed^{\ell
      -1}k}\right)^{1+\frac{1}{\ell-1}}$ constraints.

\section{The Upper Bound}

In this section we complement our lower bound by an upper bound. The
ratio of upper and lower bound will be polynomial in $k$ and $d$, but
the degree of the polynomial will depend on $\ell$.  \\

The proof of the upper bound uses the first moment method and proceeds
in two steps. First, we show that for given $n, k, d$ and $\ell$, we
can find an $\ell$-disjoint $(k,d)$-CSP $F$ over $n$ variables with
``many'' clauses. In a second step, we replace each literal $x \ne b$
in each constraint of $F$ by $x \ne b'$, where $b'$ is each time
chosen independently uniformly at random from $\{0,1,\dots,d-1\}$,
resulting in a random $\ell$-disjoint $(k,d)$-CSP $F'$.  We will show
that for the right values of $n$, $F'$ is unsatisfiable with positive
probability.\\

As long as we do not care about the values $b$ in the literals, a CSP
is basically nothing more than a hypergraph.

\begin{lemma}
  Let $\ell \leq k \leq n$. There exists an $\ell$-disjoint
  $k$-uniform hypergraph with
$$
m = \left\lceil\frac{{n \choose l}}{{k \choose l}^2}\right\rceil
$$
edges.
\label{large-hypergraph}
\end{lemma}

\begin{proof}
  We will actually prove something stronger. Let $\mathcal{S}$ be the
  set of all $k$-sets of $\{1,\dots,n\}$. We claim that any {\em
    maximal} $\ell$-disjoint subfamily $\mathcal{H} \subseteq
  \mathcal{S}$ has at least $m$ sets. Suppose $\mathcal{H} \subseteq
  \mathcal{S}$ is maximal. For $A,B \in \mathcal{S}$, we say $A$ is
  {\em incompatible} with $B$ $|A \cap B| \geq \ell$. Note that by
  this definition, $A$ is incompatible with itself.  By maximality of
  $\mathcal{H}$, each $A \in \mathcal{S}$ is incompatible with some $B
  \in \mathcal{H}$. For each $B \in \mathcal{H}$, there are at most
  $$
  {k \choose \ell}{n-\ell \choose k-\ell}
  $$
  sets $B \in \mathcal{S}$ incompatible with $A$: Each fixed
  $k-\ell$-subset of $A$ is contained in ${n-\ell \choose k-\ell}$
  subsets of $\{1,\dots,n\}$, and $A$ contains ${k \choose \ell}$ such
  $\ell$-subsets.  Hence $|\mathcal{S}| \leq {k \choose \ell}{n-\ell
    \choose k-\ell}|\mathcal{H}|$, and the claim follows after a short
  calculation.  \qed
\end{proof}

We bound $m$, the size of the $\ell$-disjoint $(k,d)$-hypergraph on
$n$ vertices, from below by a formula that will be easier to work with:
\begin{eqnarray}
  m \geq \frac{{n \choose \ell}}{{k \choose \ell}^2} & \geq &
  \left(\frac{n}{k}\right)^{\ell} \frac{\ell^\ell}{(ek)^\ell}
  = n^\ell \left(\frac{\ell}{ek^2}\right)^\ell \ .
\label{formula-m}
\end{eqnarray}

We can obtain a $(k,d)$-CSP over variable set $V=\{x_1,\dots,x_n\}$
from a $k$-uniform hypergraph over vertex set $\{v_1,\dots,v_n\}$ by
simply replacing each edge $\{v_1,v_2,\dots,v_k\}$ by a constraint
$\{x_1 \ne b_1, \dots, x_k \ne b_k\}$, where we sample each $b_i$
independently and uniformly at random from $\{0,\dots,d-1\}$. We
obtain a random CSP $F$. Any fixed assignment $\alpha$ has a chance of
$d^{-k}$ to satisfy a random constraint, and each random constraints
is chosen independently. Hence $\alpha$ satisfies $F$ with probability
$\left(1-d^{-k}\right)^m$, where $m=|F|$ is the number of constraints.
The expected number of satisfying assignments of $F$ is

\begin{eqnarray}
\sum_{\alpha:V \rightarrow \{0,\dots,d-1\}} \Pr[\alpha
\textnormal{ satisfies } F] = d^n \left(1-d^{-k}\right)^m
< e^{\ln(d)n - d^{-k}m} \ .
\label{number-sat}
\end{eqnarray}

If we can choose $n$ and $m$ such that the latter term is $\leq 1$,
then with positive probability, $F$ is not satisfiable. We re-write
this condition:
\begin{eqnarray*}
\ln(d)n - d^{-k}m & \leq & 0 \leftrightarrow \\
m \geq ln(d) n d^k
\end{eqnarray*}
Combining this with (\ref{formula-m}), we see that it suffices
to choose $n$ such that
$$
n^{\ell}n^{-1} \geq \ln(d) \left(\frac{ek^2}{\ell}\right)^\ell d^k \ ,
$$
and we choose
$$
n := \left\lceil \left(\frac{ek^2}{\ell}\right)^{\frac{\ell}{\ell-1}}
    \left(\ln(d)d^k\right)^{\frac{1}{\ell-1}} \right\rceil \ .
$$
Hence there is some constant $c$
such that
$$
m = \left\lceil n^\ell \left(\frac{\ell}{ek^2}\right)^\ell \right\rceil
\leq c \left(\frac{ek^2}{\ell}\right)^{\frac{\ell^2}{\ell-1}-\ell}
\left(\ln(d)d^k\right)^{\frac{\ell}{\ell-1}}  
= c\left(ek^2\ell^{-1}\ln(d)d^k\right)^{1+\frac{1}{\ell -1}} \ .
$$

With these values of $n$ and $m$, the rightmost term in
(\ref{number-sat}) is $\leq 1$, and thus with positive probability,
the random $(k,d)$-CSP $F$ has $0$ satisfying assignments.
This finishes the proof of Theorem~\ref{main-result}.$\hfill\Box$\\

\section{Conclusions and Open Problems}

We determined the value of $m_\ell(k,d)$ up to a factor that is,
for constant $\ell$, polynomial in $k$ and $d$. Can one eliminate
the exponential factor $d^{-\ell +1}$ in the lower bound? \\

Further, we do not have any good {\em explicit} construction of
unsatisfiable linear $k$-CNF formulas. Can one derandomize our
randomized construction? Our lower bound suffers from a similar
problem: Given an $\ell$-disjoint $(k,d)$ -CSP formula $F$ with $\leq
\left(\frac{d^k}{ed^{\ell-1}k}\right)^{\frac{1}{\ell -1}}$ frequent
variables, we know that $F$ is satisfiable, but we do not know how to
find a satisfying assignment in polynomial time.\\

Last, can one obtain any good lower bound on $m_\ell(k,d)$ that does
not use the Lov\'asz Local Lemma?

\bibliographystyle{splncs}
\bibliography{refs}

\end{document}